# Unsupervised Hyperspectral Stimulated Raman Microscopy Image Enhancement: Denoising and Segmentation via One-Shot Deep Learning

Pedram Abdolghader[1,2], Andrew Ridsdale[2], Tassos Grammatikopoulos[3], Gavin Resch[1], François Légaré[4], Albert Stolow[1,2,5,6,7], Adrian F. Pegoraro[1,2,9], Isaac Tamblyn[1,2,8,10]

[1] *Department of Physics, University of Ottawa, Ottawa, ON, K1N 6N5, Canada*
[2] *Security and Disruptive Technologies, National Research Council Canada, Ottawa, ON, K1A 0R6, Canada*
[3] *SGS Canada Inc, Lakefield, ON, Canada*
[4] *Institut National de la Recherche Scientifique, Centre EMT, Varennes, QC, J3X1S2, Canada*
[5] *NRC-uOttawa Joint Centre for Extreme Photonics, Ottawa ON K1N 6N5 Canada*
[6] *Max-Planck-uOttawa Centre for Extreme and Quantum Photonics, Ottawa ON K1N 6N5, Canada*
[7] *Department of Chemistry, University of Ottawa, Ottawa, ON, K1N 6N5, Canada*
[8] *Vector Institute for Artificial Intelligence, Toronto, Ontario, Canada*
[9] apegorar@uottawa.ca
[10] isaac.tamblyn@uottawa.ca

**Abstract:** Hyperspectral stimulated Raman scattering (SRS) microscopy is a label-free technique for biomedical and mineralogical imaging which can suffer from low signal-to-noise ratios. Here we demonstrate the use of an unsupervised deep learning neural network for rapid and automatic denoising of SRS images: UHRED (Unsupervised Hyperspectral Resolution Enhancement and Denoising). UHRED is capable of "one-shot" learning; only one hyperspectral image is needed, with no requirements for training on previously labelled datasets or images. Furthermore, by applying a *k*-means clustering algorithm to the processed data, we demonstrate automatic, unsupervised image segmentation, yielding, without prior knowledge of the sample, intuitive chemical species maps, as shown here for a lithium ore sample.

## 1. Introduction

Coherent Raman Microscopy (CRM) is a powerful nonlinear optical imaging technique based on contrast via Raman active molecular vibrations. CRM offers rapid, sensitive, chemical-specific and label-free 3D sectioning of samples and has been applied to fields ranging from biology, to medicine, to mineralogy [1-6]. Although acquisition speeds at up to video-rate have been developed for imaging at a single (fixed) Raman shift, the spectral overlap of Raman bands and the presence of other non-resonant optical signals can challenge the attribution of image intensity to specific chemical species. Therefore, multiplex or hyperspectral Raman imaging modalities [4, 7-10] were developed to address this issue. Broadband hyperspectral Raman imaging often takes advantage of synchronized fs oscillators, optical parametric oscillators (OPO's), or supercontinuum generation in Photonic Crystal Fibres (PCFs). Spectral focusing [10-19] permits high Raman spectral resolution from these broadband sources. These advances resulted in improved CRM imaging and, of particular relevance here, much larger data volumes. Strategies proposed to analyze hyperspectral datasets include multivariable curve resolution

[20-21], independent component analysis [22], vertex component analysis [23] and spectral phasor analysis [24]. In CRM, these methods assign a meaningful label to each pixel based on Raman spectroscopy, helping to classify materials in the sample, thus allowing the image to be segmented. Unfortunately, a common issue afflicting CRM is low signal-to-noise ratios (SNRs) which can limit the ability to segment a dataset. Low SNR can originate from, as examples, laser source noise, low concentrations of target molecules in the sample, and/or scattering losses in deep-tissue imaging [25-26]. Furthermore, in some CRM imaging applications (e.g., *in vivo*), fast imaging and/or low input laser power is required to prevent sample photo-damage, typically resulting in low contrast (low SNR) images. Low SNR images often additionally suffer from poorly resolved spectral features. To date, various denoising algorithms have been applied to image enhancement. However, these typically require either prior knowledge of the noise source or averaging multiple images of the same field of view (under better observation conditions), potentially resulting in reduced spatial-spectral resolution [27-28].

Machine learning has emerged as a powerful and general tool for scientific data analysis, having demonstrated remarkable performance in imaging applications such as super-resolution microscopy and cancer detection, amongst others [29-36]. Applications of machine learning in CRM have resulted in, for example, advances in lung cancer diagnosis [35], the analysis of expressed human meibum [37] and sample classification [38]. The well-known U-net architecture [39] is one example of deep learning applied to denoising SRS microscopy images [40] and was applied to denoising Raman spectra in a supervised manner [41]. The U-net architecture has been further extended to make use of both spatial and spectral correlations in hyperspectral CRM imaging to predict fluorescence-labelled images of cells based solely on label-free CRM spectra [42]. Deep learning is a subset of machine learning based on hierarchical and multi-layer artificial neural networks. A powerful customized approach named DeepChem was applied to SRS microscopy for material classification and fast imaging [43]. DeepChem overcomes the need for spectrally resolved Raman spectra in order to segment images; however, the degree to which species can be differentiated is limited and requires spectrally resolved, labelled training data as an initial input. Indeed, previous applications of deep learning to CRM imaging were based on supervised learning approaches. That is, they require labelled training data, either in the form of high SNR data for image and spectral denoising or spectrally resolved data in the case of DeepChem. While obtaining such data is possible in some cases - for delicate or rare biological samples, or for tracking transient events (requiring high frame-rate imaging) - acquiring high SNR training sets may be more challenging.

Here we present an unsupervised deep learning approach to processing CRM images which removes the need for independently acquired and labeled training data. Our approach, UHRED (Unsupervised Hyperspectral Resolution Enhancement and Denoising) can automatically, without human oversight, segment and label distinct features within a dataset. We compare this unsupervised approach with a supervised one, SHRED (Supervised Hyperspectral Resolution Enhancement and Denoising), and demonstrate that the unsupervised model compares favorably. We recorded spectrally-resolved nonlinear optical signals (termed here the hyperspectral index) at each image pixel. Importantly, the hyperspectral index is general and could represent any linear or nonlinear optical signal channels including SRS, Pump-Probe, thermal lensing and cross-phase modulation microscopy or any other signal types which depend on an input laser parameter. In the case illustrated here, the hyperspectral index is the SRS vibrational Raman spectrum. We anticipate that this "one-shot" approach will help broaden the applications of deep learning to hyperspectral optical and nonlinear optical microscopy.

## 2. Methods

### 2.1 Hyperspectral Stimulated Raman Scattering Microscopy

Our spectral focusing hyperspectral SRS imaging arrangement was reported previously [10-13]. A schematic of the experimental optical apparatus is illustrated in Fig. 1. Briefly, a fs dual output laser system (InSight DS +, Spectra-Physics, USA) produced two synchronized pulse trains at 80 MHz. The fixed wavelength 1040 nm output had a transform-limited pulse duration of 220 fs, with an average power of 1 W. The second output was tunable over the 680-1300 nm range, with a transform-limited pulse duration of approximately 120 fs at a maximum of 1.5 W average power. The 1040 nm output was amplitude modulated at 1 MHz using a Pockels cell (350-160, Conoptics, USA) driven by an external function generator (DS345, Stanford Research Systems, USA). Rapid tuning of the Raman resonant frequency was achieved via time-delay scanning of the chirped pulses. After recombination, both beams were linearly chirped by propagation through 60 cm of SF11 glass before being sent to an inverted microscope (IX-71, Olympus, Japan) equipped with galvanometer mirrors for raster scanning a 2D image. The beams were focused on the sample with a microscope objective (UPlanSapo, 20x, NA 0.75, Olympus, Japan). After the sample, the forward propagating beams were collected by a second objective acting as the condenser (LUMPlanFI/IR, 40x, NA 0.8 water immersion, Olympus, Japan). In this Stimulated Raman Loss (SRL) arrangement, the amplitude modulated Stokes beam was blocked by optical filters (BrightLine 850/310, Semrock, USA and 1064-71 NF, Iridian, Canada), whereas the transmitted Pump beam (containing the transferred modulation) was measured using a large-area photodiode (FDS10X10, Thorlabs, USA). Typically, a few mW of optical power impinged onto the photodiode which was reverse-biased at 50 V. An RF bandpass filter (#3128, KR Electronics) centred at 1 MHz frequency filtered the photodiode electrical output signal. The filtered signal was amplified by a transimpedance amplifier (DHPCA-100, Femto Messtechnik GmbH, Germany), providing the signal input to the lock-in amplifier (UHFLI, Zurich Instruments) which was synchronized with the function generator driving the Pockels cell. A lock-in time constant of 20 μs was used, and the relative phase of the lock-in amplifier was adjusted for maximum SRS signal. Data collection and galvanometer synchronization was performed using ScanImage [44].

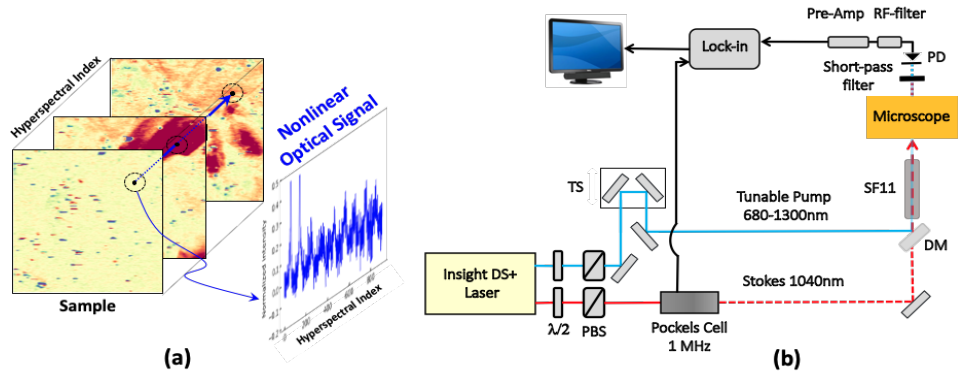

Fig. 1. Experimental implementation of hyperspectral nonlinear optical microscopy. (a) Hyperspectral image (data cube) comprised of raster-scanned 2D images (slices) as a function of a selected laser parameter (e.g. frequency, time delay, Raman shift etc.), termed here the hyperspectral index of which only three slices are shown. To illustrate the data cube, we show the variation of a particular nonlinear optical signal at a given pixel (centre of dashed circle). The blue arrow axis indicates the hyperspectral index variation from slice-to-slice. An exemplary nonlinear signal at this single pixel is plotted, in blue, as a function of its hyperspectral index. (b) Simplified implementation of hyperspectral Stimulated Raman Scattering (SRS) microscopy. Depicted are the dual output fs laser system, half wave plates (λ/2), polarizing beam splitter (PBS), Pockels cell (1 MHz), translation stage (TS), dichroic mirror (DM), highly dispersive glass rod (SF11), microscope, photodiode (PD), RF-filter (bandpass filter at 1 MHz), transimpedance pre-amplifier, and lock-in amplifier. For details, see the text.

In spectral focusing SRS, the Raman mode probed depends on the temporal overlap of the Pump and Stokes. As the temporal overlap is changed, the instantaneous excitation intensity may also vary. To account for this variation and to normalize the recorded SRS vibrational spectra, sum-frequency generation (SFG) of the instantaneous Pump and Stokes overlap as a function of their time delay was independently recorded in the microscope. In the epi-direction, a dichroic mirror (720DCXXR, Chroma, USA) directed back-reflected SFG signals through a short pass filter (750SP, Chroma, USA) to a photomultiplier tube (Hamamatsu H10723-01). KDP powder was used to generate the broadband SFG signals (data not shown) for *in situ* spectral calibration. Since SFG and SRS are each linear in the Pump and Stokes powers, SFG was used routinely to normalize the spectrally resolved SRS signals from samples.

We acquired SRS images (a data cube) where every pixel has an independently measured hyperspectral response as illustrated in Fig. 1(a). Two types of samples were used. In each, a 2D slice was 256x256 pixels with a pixel dwell time of 32 µs. The first type of sample was a heterogeneous mixture of immiscible hexadecane and distilled water. Hexadecane has a well-known Raman C-H stretch vibrational resonance at 2852 cm$^{-1}$. To record SRS spectra, the tunable Pump beam was set to 802 nm whereas the Stokes was fixed at 1040 nm. The SRS spectrum was recorded over 92 frames as the time delay between the two beams was varied. Two images of the same field of view but at different input laser powers were recorded to provide, respectively high SNR ground truth (GT) data and low SNR test data. The high SNR hyperspectral image was obtained using high input laser power (60 mW average power at the microscope input for each of Pump and Stokes). A low SNR hyperspectral image was obtained using 20 mW average laser power in each beam, thereby increasing the relative contribution of Poisson noise and reducing overall SNR, resulting in noisy Raman spectra at every pixel. Note that when comparing high SNR data to low SNR data with hyperspectral images, the comparison can be made along spatial dimensions or spectral dimensions. To calculate the spatial noise for a given image, a local mean and standard deviation (5-pixel radius neighborhood) were calculated, allowing a calculation of an effective SNR for every pixel. To calculate spectral noise, the time evolution of each pixel was compared to a reference time series and the peak signal to noise ratio (PSNR) was calculated.

A second sample type, used here to demonstrate the application of unsupervised deep learning for denoising and segmentation, was a mixture of mineralogical ores (spodumene, feldspar and quartz) from a lithium mining operation. Spodumene, $LiAlSi_2O_6$, is a lithium-bearing mineral which has become increasingly important due to the burgeoning electric vehicle industry. Feldspars are aluminosilicate minerals with the general formula $AT_4O_8$ where A is potassium (K), sodium (Na), or calcium (Ca), and T is silicon (Si) and/or aluminum (Al). Quartz, $SiO_2$, is also seen in the samples. This complex sample presented an SRS imaging challenge, revealing several different SRS bands and other nonlinear optical response peaks. Here, the Pump beam was tuned over the 929-998 nm range and both the Pump and Stokes were adjusted to 70 mW average power at the microscope input. The data acquisition scheme was as above, and the spectral scan was acquired over 909 frames.

*2.2 Neural network models*

Both deep learning approaches considered here (supervised and unsupervised) were based on the same neural network architecture (a convolutional autoencoder, shown in Fig. 2). The encoding module had four convolutional layers followed by a fully connected layer. Convolutional layers consist of an array of kernels with multiple channels that operate on the input signal. A max-pooling layer (yellow) was applied at the end of each convolutional layer, decreasing dimensionality. The Encoding module was connected to the latent space through a fully connected layer. The Decoding module had one fully connected layer and four deconvolutional layers. Each deconvolutional layer applies an up-sampling operator to its input. Both the Encoding and Decoding modules used *tanh* as a non-linear activation function which was found to outperform other activation functions tested (ReLU and leaky-ReLU). The kernel size of each convolutional layer and number of nodes in the fully connected layer determines the number of learnable parameters which must be optimized through training (details of the hyperparameters are given in the Supplementary, Fig. S1.). Depending on the range of the measured Raman shifts in each experimental sample (hexadecane vs. lithium-ore), some of the hyperparameters differed between models. However, the overall architecture (numbers and types of layers) was the same for all datasets. Hyperparameter optimization was achieved by optimizing the loss on a validation set taken within the same field of view. Full hyperparameters for each model are given in the SI along with datasets and code necessary to reproduce our results.

The difference between the supervised and unsupervised models came from the choice of input and output (target) data. For the supervised case, we used noisy, low SNR data as input and high SNR (ground truth) data as the reconstruction output target. This is a classic approach to neural network denoising [45]; the network is tasked with taking a noisy image and converting it to a 'clean' one. In our unsupervised approach, low SNR data was used both as input and as the reconstruction target output. Because the noise is random across all pixels, the ability of the autoencoder to reconstruct this high-entropy component is limited and thus only the major features of interest present across many pixels are recovered on average. In all cases, our loss was defined as the mean squared error between the neural network output and the target data. A gradient based optimizer (Adam [46]) was used to determine model parameters.

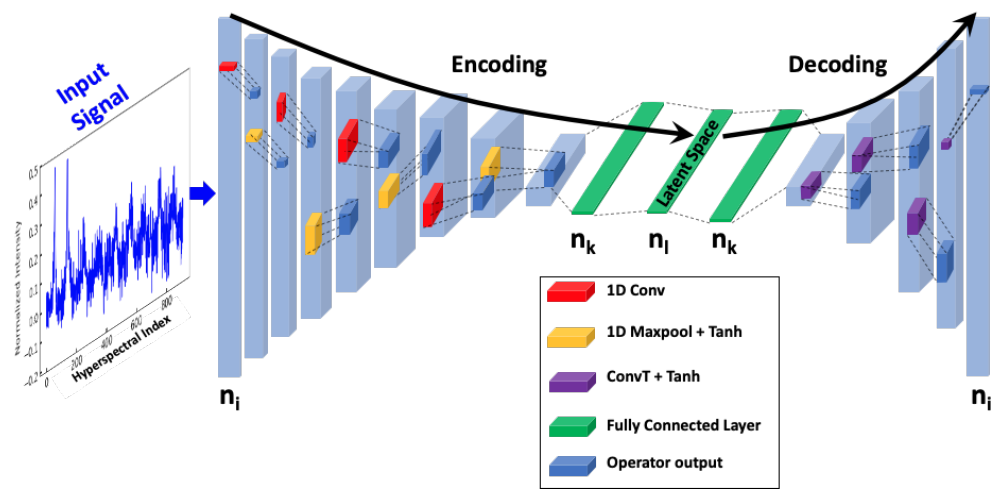

Fig. 2. The convolutional autoencoder used for our supervised and unsupervised models. The Encoding module has four convolutional layers and one fully connected layer. Each convolutional layer (red) applies a 1D kernel to its input signal, yielding an output (blue) for the next layer, to which a 1D Maxpool kernel (yellow) is then applied. This cycle is repeated until the last convolutional layer which connects to a fully connected layer (green). The latent space (green) is the bottleneck which contains the compressed representation of the input signal. It is connected by another fully connected layer (green) to a Decoding module composed of 4 transpose convolution layers. $n_i$ is the dimension of the input signal, $n_l$ is dimension of the compressed latent space. The output dimension is the same as the input, and $n_i$. We considered both supervised and unsupervised models with this architecture. Full details of the network hyperparameters are reported in the SI.

In hyperspectral imaging, each spatial pixel contains a full spectrum. A 256×256 pixel image therefore provides 65,536 1D spectra. Here, 80% of these spectra were used for training the artificial neural network model with the remaining 20% as the validation dataset. The spatial location of each pixel was not used in training and instead each pixel was treated like an independent measurement of a sample. Test data were obtained from a separate field of view within the sample which was not used for model training or hyperparameter tuning. Importantly, we achieved excellent transferability across the sample using training on only a single field of view.

We term the supervised training model SHRED (Supervised Hyperspectral Resolution Enhancement and Denoising) whereas our unsupervised model is termed UHRED (Unsupervised Hyperspectral Resolution Enhancement and Denoising). SHRED is similar to previously reported methods [35-38, 40-43]. The Pytorch framework was used to build all models and the training time was approximately 10 minutes per model, using an NVidia K80 GPU.

The second model, designed to demonstrate an unsupervised denoising ability, was also based on a 10-layer convolutional autoencoder. This model's architecture is the same as shown in Fig. 2 (details of the hyperparameters are given in the Supplementary, Fig. S2).

*2.3 Unsupervised segmentation*

In addition to denoising the spectral components (and hence composite images), a convolutional autoencoder can also be utilized for unsupervised classification (and, subsequently, image segmentation). During encoding, the dimensionality of the input data is reduced to a minimal representation (described by its projected coordinates within the latent space). Within this low dimensional space, a *k*-means clustering algorithms (hereafter, *k*-means) can be used to identify similar pixels within the image. Unlike other ML algorithms [37-38], this approach requires no labelling of samples, thus making it an unsupervised classifier. To determine the number of centroids in the *k*-means algorithm without user input we use the elbow method (see supplementary Fig. S3). The elbow method determines the cost (e.g. variances) as a function of number of clusters, *k*. As *k* increases, there will be fewer elements per cluster and the variance will decrease. The point of inflection, where the variance changes the most upon an increment in *k*, is the elbow point. In Fig. 3, we conceptually depict the unsupervised segmentation process. The hyperspectral image data is first passed through the trained Encoder (blue) and is projected into the latent space (green). A hyper-dimensional clustering (*k*-means) algorithm is applied in this space and is used to classify each image pixel into a specific cluster. As shown in Fig. 3, the trained autoencoder can assign, in an automatic and unsupervised manner, image pixels to data classes via their hyperspectral features. For the case of hyperspectral SRS, this is equivalent to creating a chemical map wherein sample constituents are uniquely assigned - in a pixel-by-pixel manner - a chemical identity based on Raman vibrational spectroscopy.

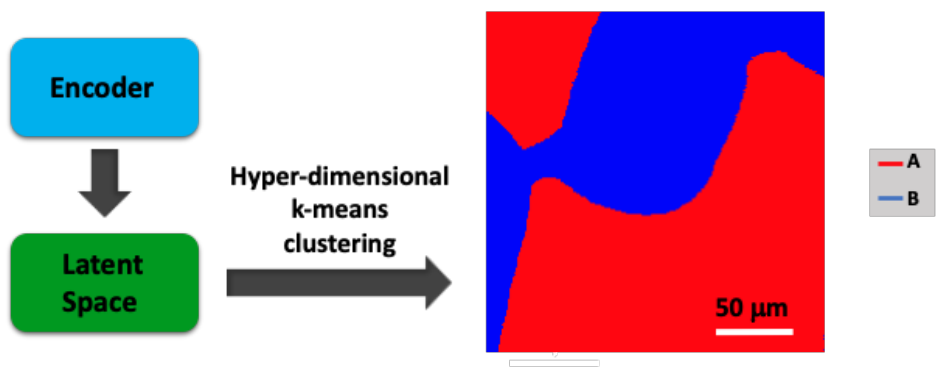

Fig. 3. Conceptual depiction of image segmentation using the trained neural net autoencoder. The hyperspectral image is passed through the trained encoder (blue). A hyper-dimensional clustering (*k*-means) algorithm is applied to the latent space (green). The optimal number of clusters, *k*, is obtained using the elbow (point of inflection) method (see Fig. S3). We demonstrate, with a pedagogical example, the application of this classification procedure to hyperspectral SRS imaging data from Fig 5(a) FOV1. The *k*-means procedure automatically and, without supervision, classifies the image components into unique chemical constituents: in other words, a chemical species map. Here red (A) is hexadecane and blue (B) is water.

## 3. Results and discussion

### 3.1 Image denoising

The trained models (both SHRED and UHRED) were used to demonstrate the denoising and image reconstruction capabilities of our autoencoder nets. An output hyperspectral image was reconstructed by feeding a low SNR hyperspectral image through the trained autoencoder nets. In Fig. 4 we show a spectral slice at 2852 cm$^{-1}$ Raman shift, which is near a Raman peak in the hexadecane. All images are normalized by the maximum pixel reading in the dataset; for the noisy Input image, this has the effect of somewhat supressing the apparent dynamic range yet allowing the noisy pixels to be resolved. As is apparent from the images, two different phases are clearly distinguishable with the hexadecane phase having a much higher signal. To compare the denoising capabilities of the different neural nets, we can measure the SNR for each phase. For the GT image (60 mW input power) shown in Fig. 4(a), the SNR in hexadecane is 31 ± 3 dB and in water is 10 ± 4 dB. The SNR of the Input image (20 mW input power) shown in Fig. 4(b) is 15 ± 2 dB in hexadecane and -8 ± 8 dB in water. For the SHRED processed image shown in Fig. 4(c), the SNR in hexadecane is 23 ± 3 dB and in water is 5 ± 3 dB, which is comparable to the UHRED processed data shown in Fig. 4(d) where the SNR in hexadecane is 23 ± 3 dB and in water is 4 ± 3 dB. Both SHRED and UHRED operate on the spectrum and do not make use of spatial information of the associated pixels; nonetheless, both SHRED and UHRED improved the SNR. We note, however, that here the same field-of-view was used for both the training (GT) and reconstructed images.

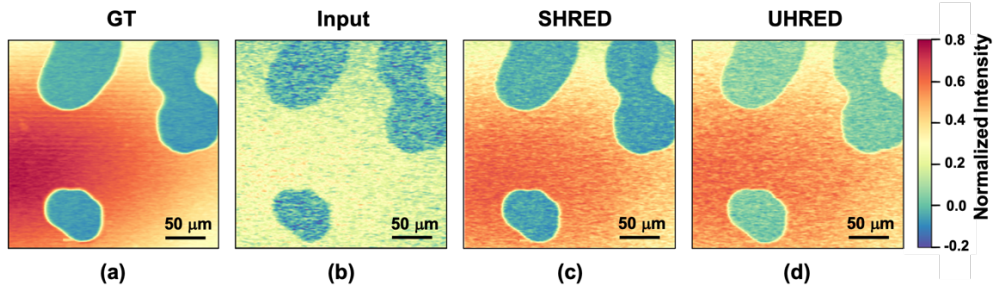

Fig. 4. SHRED (Supervised Hyperspectral Resolution Enhancement and Denoising) and UHRED (Unsupervised) methods of image enhancement for hyperspectral SRS of a binary mixture of hexadecane and water, recorded over an invariant field of view (FOV) at (a) high signal-to-noise ratio (ground truth, GT, 60mW input power), and (b) low signal-to-noise ratio (20 mW input power). In (c) and (d), we show the output of both the SHRED and UHRED models, demonstrating their denoising and contrast enhancement capabilities. For the hexadecane phase, the SNR is 31 ± 3 dB, 15 ± 2 dB, 23 ± 3 dB, and 23 ± 3 dB for the GT, Input, SHRED, and UHRED images respectively. For the water phase, the SNR is 10 ± 4 dB, -8 ± 8 dB, 5 ± 3 dB, and 4 ± 3 dB for the GT, Input, SHRED, and UHRED images respectively. All images are normalized by their maximum value.

In order to better assess our autoencoder denoising capability, we applied the model trained on one dataset (i.e. the data in Fig. 4) to the denoising of two new hyperspectral imaging datasets not previously seen by the model, shown in Fig. 5 as FOV 1 and FOV 2. While denoising was applied to the entire hyperspectral dataset, we show exemplary images taken at a Raman shift of 2852 $cm^{-1}$. In Fig. 5 (a) we show noisy input images acquired at low SNR (20 mW input power). In Fig. 5(b) and 5(c) we show, respectively, the UHRED and SHRED images reconstructed from Fig. 5(a). In Fig. 5(d) we shown the same FOV imaged at high SNR, representing the GT image (60 mW input power). Importantly, the GT image was not used here for training and is shown only to permit comparison with the UHRED and SHRED outputs. For FOV 1, the SNR in hexadecane is 15 ± 2 dB at the input, 27 ± 2 dB for UHRED, 27 ± 2 dB for SHRED, and 30 ± 3 dB for the GT and the SNR for water is -4 ± 6 dB at the input, 6 ± 7 dB for UHRED, 9 ± 3 dB for SHRED, and 18 ± 6 dB for GT. For FOV 2, the SNRs are similar with the SNR in the hexadecane being 14 ± 3 dB at the input, 25 ± 4 dB for UHRED, 30 ± 3 dB for SHRED, and 30 ± 3 dB for GT and the SNR for water being -8 ± 6 dB at the input, 8 ± 1 dB for UHRED, 10 ± 3 dB for SHRED and 10 ± 3 dB for the GT. It can again be seen that both ML approaches significantly improve image quality despite operating only on the spectral data. To further illustrate denoising performance, we plot in Fig. 5(e) the pixel intensity along the dashed line through a ~15 μm droplet within the Region of Interest (ROI) of FOV 2. It can be seen that the droplet boundaries are much more clearly defined in the UHRED and SHRED images. Indeed, using the GT data as a reference, we find that the PSNR of the noisy data is 14 dB while the PSNR using UHRED is 22 dB and the PSNR using SHRED is 25 dB.

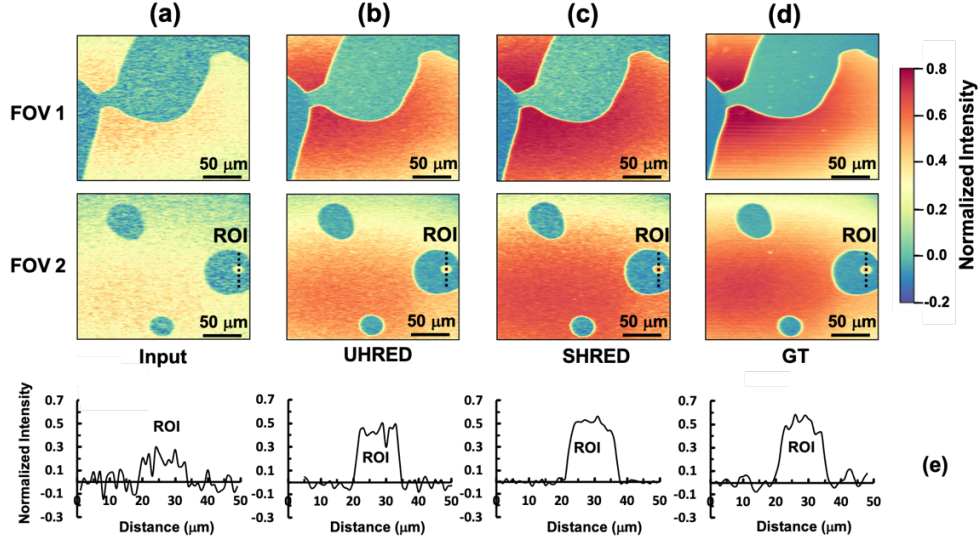

Fig. 5. Demonstration of UHRED and SHRED for SRS microscopy. Hyperspectral SRS imaging of two different fields of view, FOV 1 and FOV 2, shown in the upper and middle rows, for a binary mixture of hexadecane and water. The neural net was trained on the image dataset shown in Fig. 4 and then applied here to the FOV 1 and FOV 2 datasets which were not seen during training. Column (a) is low SNR hyperspectral images (20 mW input power). Column (b) shows the UHRED output and column (c) the SHRED output. Column (d) shows a high SNR hyperspectral image (GT 60 mW input power) which is shown here only for comparison with UHRED and SHRED. The SNR within each image depends upon the material phase. For FOV1, the SNR in the hexadecane for (a)-(d) is $15 \pm 2$ dB, $27 \pm 2$ dB, $27 \pm 2$ dB, and $30 \pm 3$ dB and the SNR in the water for (a)-(d) is $-4 \pm 6$ dB, $6 \pm 7$ dB, $9 \pm 3$ dB, and $18 \pm 6$ dB. For FOV2, the SNR in the hexadecane for (a)-(d) is $14 \pm 3$ dB, $25 \pm 4$, $30 \pm 3$ dB, and $30 \pm 3$ dB and the SNR in the water for (a)-(d) is $-8 \pm 6$ dB, $8 \pm 1$ dB, $10 \pm 3$ dB, and $10 \pm 3$ dB. The bottom row (e) shows signal line-outs along the dashed line through a ~15 μm droplet within the ROI of FOV 2; the PSNR of the noisy data is 14 dB while the PSNR of the UHRED data is 22 dB and the SHRED data PSNR is 25 dB. This comparison clearly demonstrates the image denoising capabilities of both UHRED and SHRED. All images are normalized by their maximum value.

### 3.2 Nonlinear optical hyperspectral signal reconstruction

Well-resolved Raman peaks are indispensable for classifying materials within a heterogeneous sample when using a broadband hyperspectral imaging technique such as SRS microscopy. While we have demonstrated that SHRED and UHRED effectively denoise images, they operate on the entire recorded Raman spectrum. To demonstrate that SHRED and UHRED also denoise SRS Raman spectra, we applied a previously trained model to a new, unseen dataset. In Fig. 6(d), we show a low SNR image (20 mW input power) of a hexadecane-water mixture and in Fig. 6(a) an associated Raman spectrum near the C-H stretch from a single pixel at the center of the highlighted ROI. In Fig. 6(b), we show the SHRED output Raman spectrum (red) of the same single pixel and in Fig. 6(c) the UHRED output for the same pixel. In Fig. 6(e) we show a GT Raman spectrum (green) recorded at high SNR (60 mW input power) as an independently recorded reference. To help visualize the improvement in spectral noise, we subtracted the spectrum recorded for the GT data from the Input, SHRED, and UHRED data. In Fig. 6(f-h), we show the Input residual, SHRED residual, and UHRED residual for the exemplary pixel with their respective mean squared errors (MSEs, $2.7 \times 10^{-2}$ for the Input, $1.5 \times 10^{-3}$ for SHRED, and $4.5 \times 10^{-3}$ for UHRED). To more readily compare the improvement in noise across the entire image, as well as to facilitate comparisons with improvement in image quality shown in Figs. 4 and 5, we calculate the PSNR for every pixel, using the GT as a reference. Using this comparison, it is apparent that the phase-boundary between the water and hexadecane shifted between the high SNR (60 mW input power) and low SNR (20 mW input power) recordings. Because the SHRED and UHRED outputs depend upon the low SNR input, the same pixels have low PSNR in all three datasets (see supplementary Fig. S4). Excluding pixels where the error is dominated by this boundary shift, we find that the average PSNR of the input data compared to the GT is $12 \pm 1$ dB while the average PSNR when comparing the SHRED output is $23 \pm 2$ dB. The average PSNR of the UHRED output is $24 \pm 3$ dB, demonstrating that both processing approaches reduce noise and improve hyperspectral contrast.

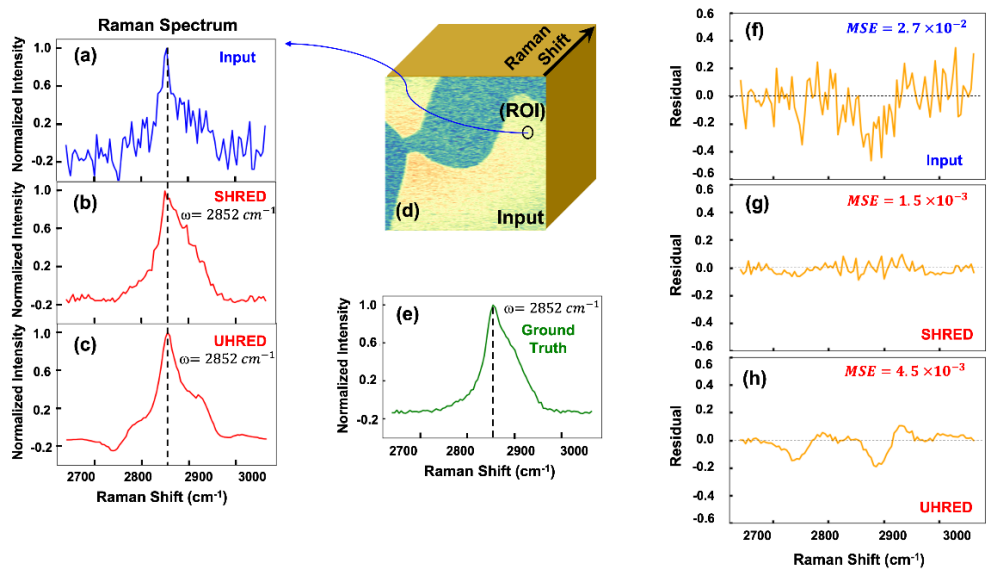

Fig. 6. SHRED and UHRED hyperspectral SRS imaging improves the C-H stretch region Raman spectrum of hexadecane from a single pixel centred within the ROI. The neural net was trained on a different FOV (data from Fig. 4) and then applied to the low SNR dataset (Input, 20 mW input power). (a) The hexadecane Raman vibrational spectrum (blue) from a single pixel centered in the ROI black circle in the low SNR image. (b) The single pixel Raman spectrum (red) as extracted by SHRED from the low SNR image for the same ROI. (c) The single pixel Raman spectrum (red) as extracted by UHRED from the low SNR image, for the same ROI. (d) Representative, low SNR image of the FOV. (e) The single pixel ground truth Raman spectrum (green) obtained via high SNR imaging (60 mW input power) of the same ROI. This ground truth spectrum was used only for comparison, not training. (f) The difference Raman spectrum (Input – GT) shows residuals with a Mean Square Error (MSE) of $2.7 \times 10^{-2}$. (g) The difference Raman spectrum (SHRED – GT) shows residuals with a MSE of $1.5 \times 10^{-3}$. (h) The difference Raman spectrum (UHRED – GT) shows residuals with a MSE of $4.5 \times 10^{-3}$.

## 3.3 Applicability of UHRED to Complex Samples

For many hyperspectral SRS datasets, Raman spectra may consist of spectral scans 'stitched together' as a laser is tuned. Furthermore, other nonlinear optical processes (e.g. cross-phase modulation, thermal lensing etc.) may contribute significant background signals to the total signal. To test the applicability of UHRED to more challenging sample types, a complex heterogeneous lithium ore sample was imaged. This sample demonstrates weak linear absorption, thus limiting the total power which may be applied. Furthermore, this sample is highly heterogeneous, strongly increasing scattering and therefore limiting the collected signal. In such cases, it may not be possible to acquire a high SNR GT dataset. We recorded hyperspectral nonlinear optical signals (i.e. the hyperspectral index) at each pixel in the lithium ore sample. Although the hyperspectral index is in principle very general, here it represents the

Pump-Stokes time delay in a spectral focussing scan, directly related to the SRS vibrational Raman spectrum. In Fig. 7 (a), we show hyperspectral imaging of Lithium ore samples at three different single pixels, indicated by the black arrow in each image. We also show the hyperspectral index plots (blue), related to SRS Raman vibrational spectra at the same pixels. These represents the 'noisy' input image data. A UHRED model was trained on an input hyperspectral data cube from a different Lithium ore sample (not shown). In Fig. 7(b), we show the UHRED output (red) at the three pixels shown in the raw input data in (a). It can be seen that UHRED significantly improves the SNR of the hyperspectral index at all pixels simultaneously, without any loss of spectral resolution. Importantly, UHRED differs from what may be achieved by other denoising methods such as a moving average smooth, as the latter reduces both spectral resolution and peak contrast. To directly compare UHRED with smoothing, we applied a moving average filter with a kernel size of 10 to every pixel in the image. The moving average spectra (m_avg) are shown in Fig. 7(a) in red, superimposed on each (blue) hyperspectral scan. It can be seen that smoothing reduces both spectral resolution and peak contrast when compared to the UHRED spectra in (b). As an aside, we note that the few green pixels seen in the images are due to detector saturation: these intense but very localized signals originate from other (non-SRS) modulation transfer signals such as transient absorption, thermal lensing, or other processes, likely due to strongly absorbing semiconductor materials (e.g. pyrite) which appear sparsely in these mineral samples.

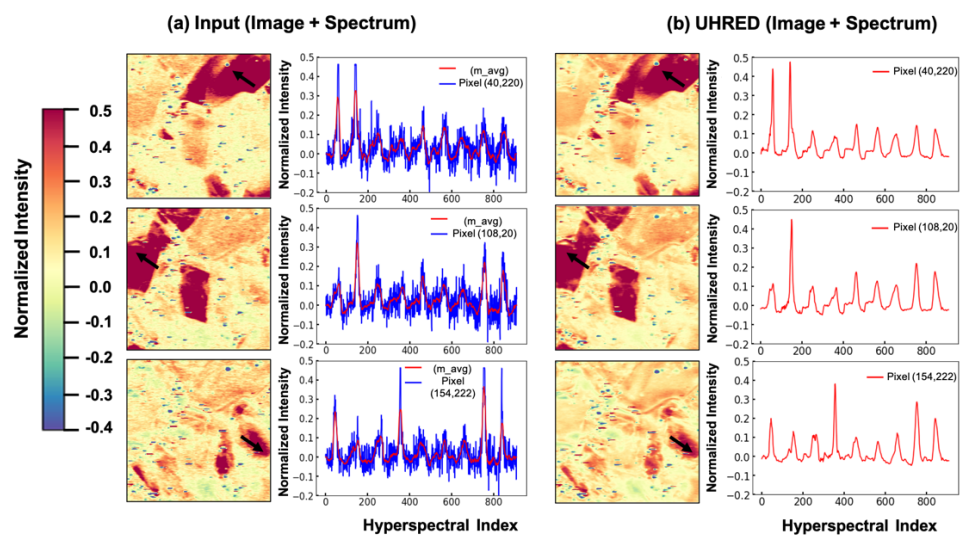

Fig. 7. Hyperspectral SRS imaging of three different fields of view in Lithium ore samples. (a) A hyperspectral image is shown along with a hyperspectral index scan (blue) for the three different pixels indicated by the black arrows in the three panels (top, middle, bottom). Application of a moving average filter with a 10 point kernel is also shown (red) superimposed on each (blue) hyperspectral scan. (b) The output of the UHRED autoencoder applied to the hyperspectral image data of (a). The UHRED output hyperspectral index scans (red) for the same three pixels from (a) are shown (top, middle, bottom). UHRED improves the hyperspectral index SNR at all pixels simultaneously, without the loss of spectral resolution or peak contrast seen in the moving average filter.

As highlighted above, it is possible to use clustering algorithms which operate in the latent space of the encoder to segment an image. Using the elbow method to determine the appropriate number of clusters (see supplementary Fig. S3(b)), we applied a *k*-means clustering algorithm to automatically segment the lithium ore dataset (Fig. 8 (top)). The majority of the sample comprised of quartz ($SiO_2$, red), feldspars ($(K, Na, Ca)_1(Si, Al)_4O_8$, yellow) and spodumene

(LiAlSi$_2$O$_6$, blue). After segmentation, it is possible to convert the hyperspectral index to SRS Raman shift, permitting direct comparison with the known Raman spectra of these mineral compounds. The randomly scattered green pixels are non-SRS modulation transfer signals saturated at the detector due to strong absorption, discussed above, and are automatically treated by *k*-means segmentation but not discussed further here. Below, exemplary single pixel Raman spectra extracted by UHRED are plotted for each constituent and compared to the reference [47] Raman spectra (black dashed lines) of these compounds. It is encouraging that the Raman spectra extracted from the model, in a completely unsupervised and automated manner, are in reasonable agreement with those of the reference compounds. This suggests that the UHRED + *k*-means method presented here is well suited to unsupervised image segmentation. We believe that this will be a useful tool in the routine generation of chemical species maps from imaging data.

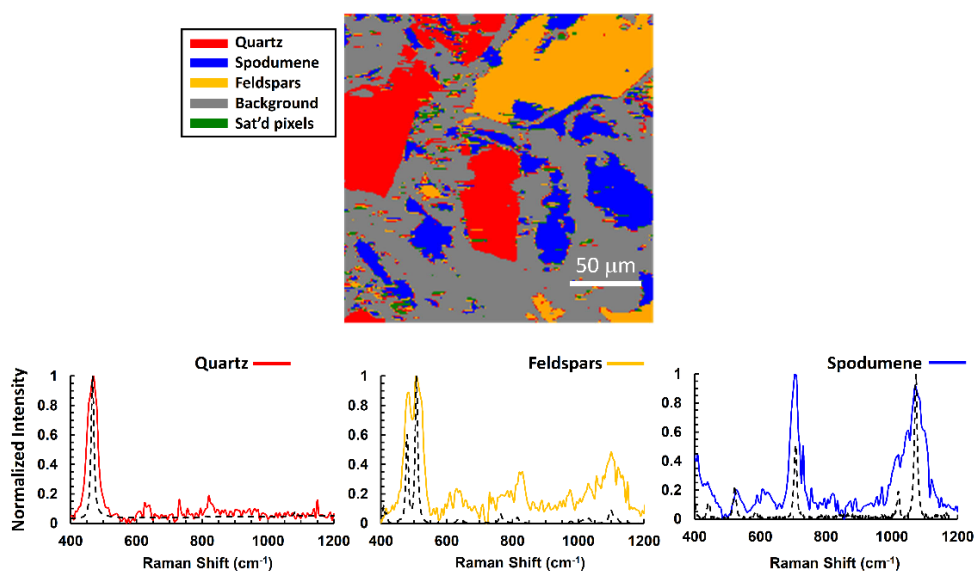

Fig. 8. Automated image segmentation of a complex Lithium ore sample (top) comprised of quartz (SiO$_2$, red), feldspars ((K, Na, Ca)$_1$(Si, Al)$_4$O$_8$, yellow) and the Lithium-bearing ore spodumene (LiAlSi$_2$O$_6$, blue). UHRED followed by an unsupervised, automated *k*-means algorithm applied in the latent space, classified constituents based on their hyperspectral index responses (see Figure 8). Here, the hyperspectral index is converted to SRS Raman shift to permit direct comparison with the known Raman spectra of these mineral compounds. Representative single pixel Raman spectra (below) extracted by UHRED + *k*-means are plotted for each constituent and compared to the reference Raman spectra (black dashed lines) of these compounds. The general agreement between the extracted Raman spectra and those of the reference compounds permits the UHRED + *k*-means method to segment the image into the chemical species map shown above. For details, see the text.

## 4. Limitations

One of the primary advantages of SRS microscopy is the ability to quantitatively record Raman spectra. In the presence of noise, the ability to ascribe variations in signal strictly to changes in the sample is reduced and may no longer be possible. The SHRED and UHRED models developed here reduce the noise but were not designed to be strictly quantitative; indeed, the spectra recovered are not perfect matches to the GT data. Despite this, relative mapping of intensity, and hence concentration, may be preserved. For example, in Fig. 7, the spatial variability of some signals in the recovered data is reflective of the variation in the sample. The degree to which relative intensities of individual pixels are reflective of the underlying chemical concentration likely depends on the amount of noise at the input in a non-trivial manner.

A related concern to recovering a quantitative map of different species is the degree to which overlapping species can be separated and classified. For example, an ore may be an amalgam of different materials, or in biological imaging, lipid and protein groups may exist within the same imaging volume. Under such circumstances, the use of *k*-means clustering as currently implemented could lead to misleading results since pixels are assigned as belonging to one group or another. However, alternative clustering and chemometric approaches could be utilized instead, either in the latent space or using the denoised output data. Additionally, it may be possible to incorporate spatial relationships between pixels, as has been recently demonstrated [42], to improve the speciation.

Both UHRED and SHRED required a preprocessing step to remove saturated pixels. As noted above, a few pixels in the raw datasets were saturated during data acquisition. To train the models, saturated pixels were replaced by the average value of a randomly chosen subset of nearby pixels. In the current implementation, saturated pixels were identified manually but this process could be automated in future implementations. As the preprocessing phase can be eliminated when a clean dataset is supplied, our technique remains unsupervised.

As was illustrated in the discussion of Fig. 6 and by supplementary Fig. S4, for microscopic samples it is possible for the sample to shift between subsequent scans. For training the SHRED net, such a shift could introduce errors because the noisy input data and the GT data no longer agree. It is worth noting, however, that this limitation does not affect UHRED because the input and output training data are identical and must agree in terms of the material phase present.

Looking at the sample residuals in Figs. 6(f) and 6(g), as well as the reconstructions in Figs. 6(b) and 6(c), it is not apparent if the differences in the recovered spectra are systematic or random. We have found that the errors are random with different pixels showing different shifts in the spectrum. Nonetheless, it raises an interesting question as to whether error contributions from the "tails" of the data could be minimized. Different regularization schemes could be tested which could minimize overfitting. Additionally, it may be possible to supplement the models with additional information. For example, if knowledge about laser intensity for a given Raman shift is known from another nonlinear optical processes, such as SFG imaging, this could be incorporated.

## 5. Conclusions

We applied unsupervised (UHRED) and supervised (SHRED) deep learning techniques to hyperspectral images in order to achieve contrast enhancement and segmentation of chemically distinct species within a sample. The hyperspectral index discussed here is quite general and can represent any laser parameter-based index which is varied. The hyperspectral index scan could therefore represent, as examples, spectral data from SRS, CARS, linear or nonlinear fluorescence, Pump-Probe, Thermal Lensing or Cross-Phase Modulation microscopies. In the examples presented here, the hyperspectral index is the SRS Raman vibrational spectrum obtained in a spectral focussing SRS arrangement. UHRED demonstrably enhances hyperspectral SRS contrast in low SNR images. Furthermore, the extracted single pixel Raman spectra were in good agreement with both the SHRED and ground truth spectra (the latter recorded at high SNR). Importantly, implementing the *k*-means clustering algorithm to the latent space provided us with an automated, unsupervised image segmentation procedure. This corresponds to an intuitive chemical species map which is useful in many fields of science and technology. UHRED can be further generalized to include spatial properties of the sample, including birefringence, and to incorporate multimodal signals such as harmonic generation, fluorescence, thermal lensing etc. As computational power continues to increase, we expect that real-time automated, unsupervised image denoising and material identification will become available to researchers worldwide.

**Acknowledgements.** We thank SGS Canada for providing spodumene mineral samples and Rune Lausten (NRC), Doug Moffatt (NRC), Kevin Ryzcko (uOttawa), Evan Thomas (uOttawa), for helpful discussions.

**Funding:** Natural Sciences and Engineering Research Council of Canada (Discovery, CRD); National Research Council Canada (AI4D Program); SGS Canada Inc.; NRC-uOttawa Joint Centre for Extreme Photonics; Max Planck - uOttawa Centre for Extreme and Quantum Photonics.

**Disclosures.** The authors declare that there are no conflicts of interest related to this article.

**Data Availability.** All data used in this study is available at Ref. [48]. All algorithms and source code are available at Ref. [49].

**Supplemental Document.** See Supplement 1 for supporting content.

# Unsupervised Hyperspectral Image Enhancement, Segmentation and De-Noising in Stimulated Raman Microscopy: supplemental document

## 1. First model

A 10-layer convolutional autoencoder trained in order to denoise the hyperspectral SRS microscopy images. Convolutional layers in the encoder modules compressed the dataset dimension and mapped them to the latent space. The latent space can be considered a feature vector where applying deconvolutional layers (CovT) on it leads to reconstructing the input dataset. Figure S1 demonstrates the first model architecture and its parameters. The dataset that was fed into this model was spectra from a mixture of hexadecane and water. Input dataset shape was 65536 signals each with a length of 92 points and one channel.

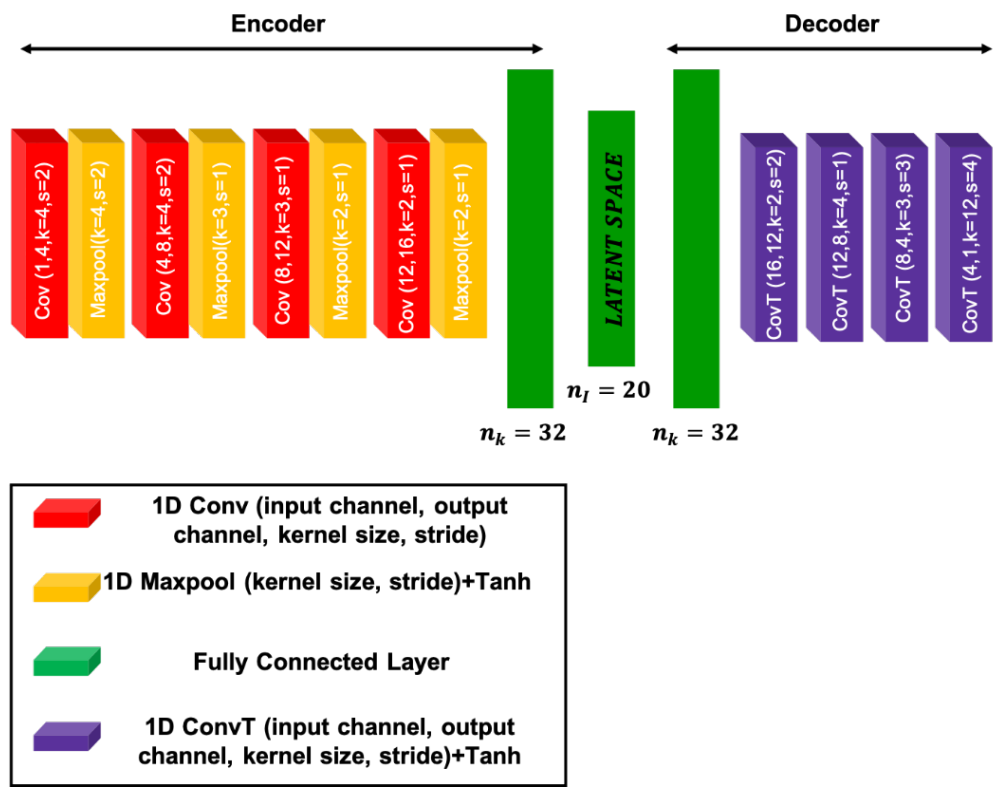

Fig. S.1. First model architecture and its parameters.

## 2. Second model

Figure S2 demonstrates the second model architecture and its parameters. The dataset fed into this model was spectra from a mixture of mineral ores. Input dataset shape was 65536 signals each with a length of 909 points and one channel.

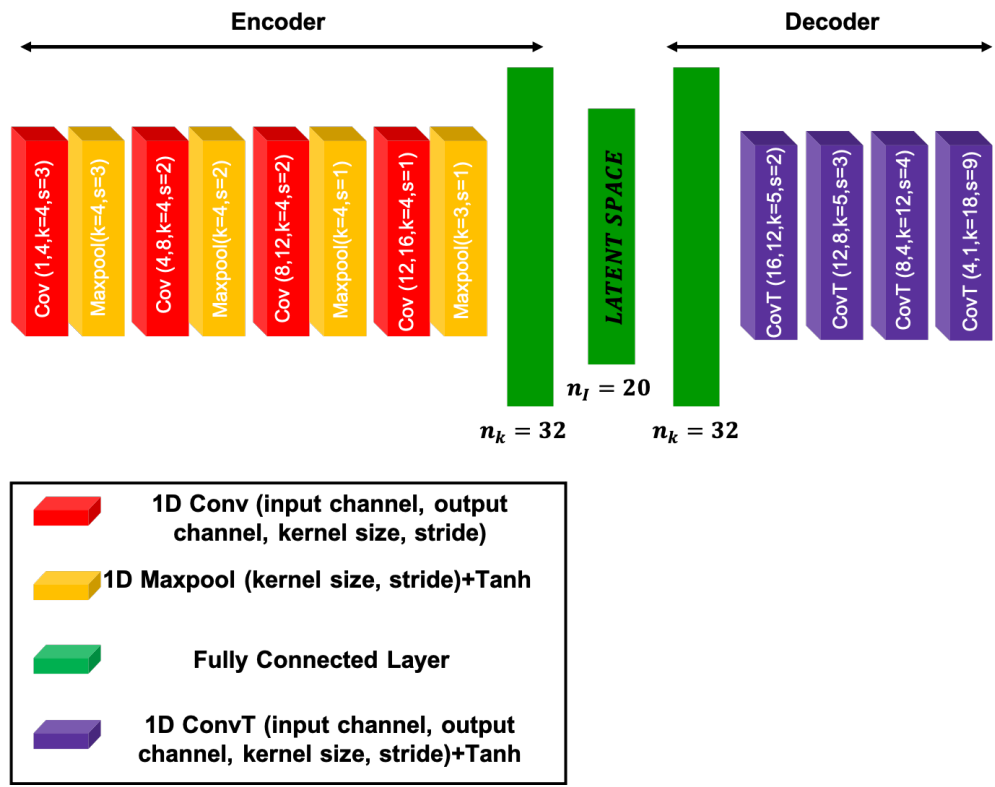

Fig. S2. Second model architecture and its parameters.

## 3. Elbow method in order to determine the best number of clusters in the latent space

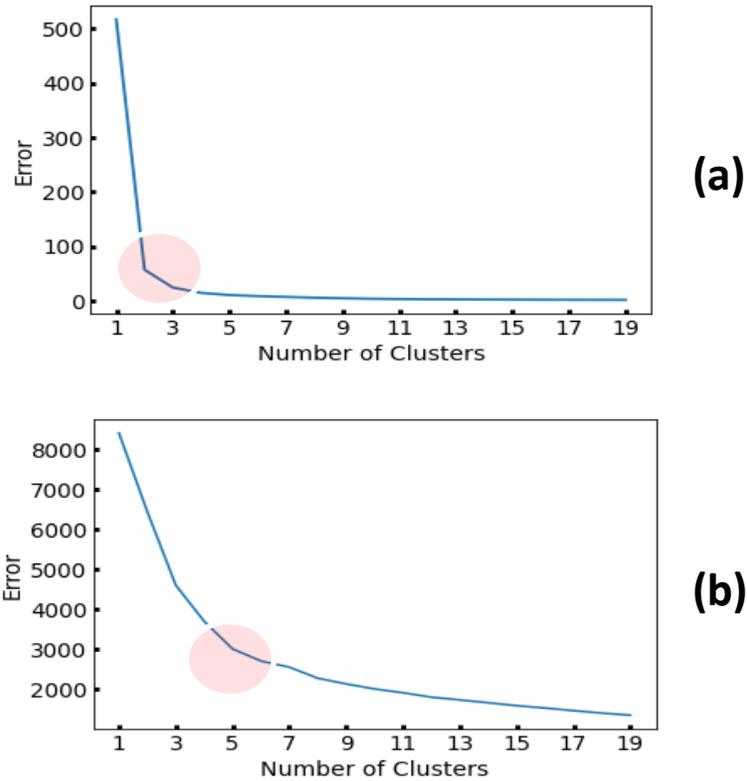

Fig. S3. Applying elbow method on the latent space (with *k*-mean algorithm) to determine the best number of clusters in latent space. (a): On hexadecane and water sample, (b) on mineral ore sample.

## 4. Spatial distribution of error showing the effect of a moving boundary between the ground truth and the noisy input image

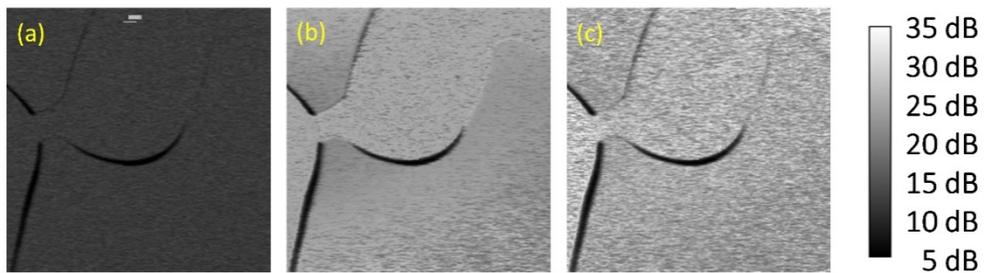

Fig. S4. The spatial distribution of the PSNR for FOV1 for the data presented in Fig. 5 for (a) Input-GT, (b) SHRED-GT, and (c) UHRED-GT. It is apparent that there was a shift in the boundary between the hexadecane and water phases which affects all comparisons equally.

**Data Availability.** All data used in this study is available at Ref. [1]. All algorithms and source code are available at Ref. [2].